# Orbital Angular Momentum in Noncollinear Second Harmonic Generation by off-axis vortex beams

**Fabio Antonio Bovino,[1*], Matteo Braccini[2], Maurizio Giardina [1], Concita Sibilia [2]**

[1] *Quantum Optics Lab, Selex S.*I.*, Via Puccini 2, 16154 Genova, Italy*
[2]*Dipartimento di Scienze di Base e Applicate per L'Ingegneria , "Sapienza" Università di Roma, Via Scarpa 16, 00161 Roma, Italy*
[*]*fbovino@selex-si.com*

**Abstract:** We experimentally study the behavior of orbital angular momentum (OAM) of light in a noncollinear second harmonic generation (SHG) process. The experiment is performed by using a type I BBO crystal under phase matching conditions with femtosecond pumping fields at 830 nm. Two specular off-axis vortex beams carrying fractional orbital angular momentum at the fundamental frequency (FF) are used. We analyze the behavior of the OAM of the SH signal when the optical vortex of each input field at the FF is displaced from the beam's axis. We obtain different spatial configurations of the SH field, always carrying the same zero angular momentum.

## 1. Introduction

Optical vortices (OVs) in light beams are tightly bound to phase dislocations (or singularities): due to the continuous spatial nature of a field, the presence of these defects implies the vanishing of the field's amplitude in the singularity; in some types of dislocations the phase circulates around the singularity and creates a vortex [1]. Nye and Berry [2] used the term "phase dislocation" to define the locus of the zero amplitude of a field. Similarly to crystallography, phase dislocations can be classified in edge, screw and mixed screw-edge [3]. The interest in OVs raised because fields in which they are included show a helical wave-front structure - developing around the screw dislocation line – entailing the presence of an orbital angular momentum (OAM). This is an important feature which can be exploited in several applications, from optical tweezers, to the generation of N–dimensional quantum states (quNits) hence to a large potential information capacity.

A screw wave dislocation can be defined by means of the integer topological charge Q which represents the winding number of the phase and can be found by means of a circulation integral around the dislocation line:

$$Q = \frac{1}{2\pi} \oint df \qquad (1)$$

where $f$ is the phase of the field.

As mentioned above, beams carrying optical vortices are characterized by an orbital angular momentum per photon of $\ell\hbar$ [4, 5]. There are several ways of generating optical vortices beams and these include cylindrical lenses [5], computer generated holograms (fork holograms) [6], spatial light modulators (SLMs) and spiral phase plates (SPPs) [7,8]. The latter are transparent optical devices which force a field, collinear with the optical axis, to experience an azimuthally dependent optical path; therefore they can be treated by means of a transmission function exp($il\theta$) where $\theta$ is the azimuthal polar coordinate and $l$ is the (integer) phase's winding number or topological charge: when a Gaussian beam hits a SPP, the outgoing field can be modeled by Laguerre – Gaussian (LG) modes or by a superposition of them:

$$u(\rho,\theta) = \sum_{l=-\infty}^{\infty}\sum_{p=0}^{\infty} C_{lp}\psi_{lp}(\rho,\theta) \qquad (2)$$

Where $\psi_{lp}$ are the LG modes, $C_{lp}$ are the corresponding coefficients, $l$ is the azimuthal index, bound to the topological charge, and $p$ is the radial index which denotes the number of nodal rings developing from the beam's axis. Since it was shown that LG modes posses an OAM per photon in ℏ units equal to their azimuthal index $l$ [5], the modal decomposition is particularly useful for calculating the OAM carried by a beam.

Of course it is possible to obtain beams with non integer OAM by enclosing in the beam a mixed screw-edge dislocation [8] or, as it will be shown later, by displacing the SPP with respect to the incident beam's axis [9, 10].

Among nonlinear optical interactions, second harmonic and parametric processes concerning fields carrying optical vortices attracted interest because they grant the possibility of manipulating the dynamics of the phase singularities [11-17]. This means that, together with the frequency doubling, the SH generated field shows phase defects correlated to the ones present in the pump beam. The properties of OAM in nonlinear interactions of OV beams were widely investigated in collinear SH processes where an *on-axis vortex pump beam* impinged on quadratic nonlinear crystal [11-12]. In reference [11] collinear SHG was studied using LG beams as pump field and it was shown that the SH field generated from a $LG_{n0}$ (n is an integer) doubles its OAM, thus verifying the momentum conservation. However *there are no studies on noncollinear SHG involving fractional OAM beams* and only little attention was placed on parametric processes with fields carrying off-axis vortices [13].

In this paper we present an experiment of noncollinear second harmonic generation where two femtosecond input beams, at the fundamental frequency of 830 nm, carrying a fractional OAM, impinge on a type I nonlinear crystal of BBO [18]. The non integer angular momentum is obtained by means of a fractional SPP, which impresses on the incident field a mixed screw edge dislocation; the latter is then displaced from the beam's axis. The "specular" configuration of the pumping beams, i.e. each beam carries the same OAM but with opposite sign, allows to obtain a SH field with different patterns all having a total OAM equal to zero.

## 2. Half integer spiral phase plate

It is known that light beams can carry angular momentum (AM). This can be split in two independent components, the spin angular momentum (SAM), bound to the polarization, and the orbital angular momentum (OAM) generated by the tangential component of the Poynting vector, thus dependent on the spatial field distribution. In terms of AM per photon, SAM gives a contribution of ±ℏ if the field is circularly polarized (0 otherwise). In our analysis we consider linearly polarized fields in order to neglect the spin contribution.

Since we are considering paraxial beams, which means that OAM must be constant during propagation, we can refer to the transverse plane at any fixed $z$ [4]. In order to calculate the intrinsic OAM, i.e. not dependent on the choice of the axis, a suitable reference system which guarantees a zero transverse momentum has to be chosen [19]. We are dealing with off-axis OVs thus the choice of a z axis coincident with the propagation axis is not so obvious: the right choice should be fixing the origin in the center of mass of the vortex beam; however the vorticity of the field is introduced by the SPP on the input Gaussian beam, hence it is possible to chose the *z*-axis parallel to the propagation axis of the beam and to place the origin at the center of the beam in the near field [21].

In this way, following a general treatment, the angular momentum density is defined as $L = \mathbf{r} \times \mathbf{P}$, being **P** the linear momentum density of the beam. If we consider a linearly polarized field propagating along *z* with a transverse distribution $u(r,\theta,z)$

$$E(r,\theta,z) = u(r,\theta,z) e^{ikz - i\omega t} \qquad (3)$$

we have for the linear momentum density the following expression:

$$\mathbf{P} = \frac{i\varepsilon_0}{\omega}\left(u\nabla_\perp u^* - u^*\nabla_\perp u\right) + 2\frac{k\varepsilon_0}{\omega}|u|^2 \mathbf{z} \qquad (4)$$

It can be seen that the *z*-component of *L* depends only on the azimuthal component of the linear momentum vector [20]. Choosing the transverse plane at *z*=0, by integrating the momentum density *L* over the transverse plane and dividing by the field energy per unit length, the OAM per photon – in ℏ units can be obtained:

$$\ell = i\frac{\int u^*(r,\theta)\frac{\partial u(r,\theta)}{\partial \theta} r\, dr\, d\theta}{\int |u(r,\theta)|^2 r\, dr\, d\theta} \qquad (5)$$

We consider a situation where a Gaussian beam hits the center of a SPP which impresses on the field an OAM per photon of $\ell=1/2$ (in ℏ units); the beam's propagation axis coincides with the optical axis of the device. As pointed out before, we deal with a mixed screw-edge dislocation. It is possible to represent the field produced by the SPP via the decomposition of Eq. (2). Substituting Eq. (2) in Eq. (5) a simple expression of the OAM, calculated with respect to the *z*-axis, can be obtained:

$$\ell = \sum_{l=-\infty}^{\infty}\sum_{p=0}^{\infty} l\,|C_{lp}|^2 \qquad (6)$$

Oemrawsingh *et al.* [21] observed that displacing a fork hologram from the beam's axis would result in a decrement of the OAM of the field. In a very similar fashion the OAM can be tuned by moving a spiral phase plate off the axis: the momentum obtained from the fractional SPP can be varied between 1/2 and 0 through the displacement of the SPP itself. The output momentum thus results smaller, following the Gaussian law presented in ref. [21] $Q*\exp(-2\, r_v^2/w_0^2)$, where Q is the topological charge impressed by the device, $r_v$ is the displacement and $w_0$ the spot size of the incident beam. In this way the OAM can be tuned to a well defined value. The decompositions of Eqs. (2) and (6) are still valid to model an off-axis dislocation and to evaluate its OAM.

## 3. Nonlinear interaction

*3.1 Experiment*

We consider a noncollinear SHG scheme in which two fractional helical beams generated by a SPP impinge with opposite OAMs on a nonlinear crystal (type I BBO) with small angles of incidence. A SH signal which carries information on the input phase's dislocations is generated. We followed the evolution of the SH field's spatial distribution as the spiral phase plate is displaced from the beam's axis and evaluated the OAM of the SH generated beam. Subsequently we compared the experimental results with the numerical ones obtained by using the modal expansion of Eq. 2.

Figure 1 shows the experimental set-up for the noncollinear SHG interaction, together with the one used for displacing the SPP (inset). In our experiment the output of a mode-locked femtosecond Ti:Sapphire laser system tuned at $\lambda$=830 nm (76 MHz repetition rate, 130 fs pulse width), collimated and spatially filtered by a 50 μm pinhole, hits a spiral phase plate, with a spot-size of about 500 μm. The distance between the lens and the SPP is 50 cm , so as the distance from the le lens to the BBO crystal. The spiral phase plate [22] was designed to operate at a wavelength of 830 nm introducing a fractional topological charge, thus an OAM per photon in ℏ units, of ½. The center of the SPP is dominated by the finite size of the height anomaly of the order of 50 microns, the SPP diameter is about 8 mm.

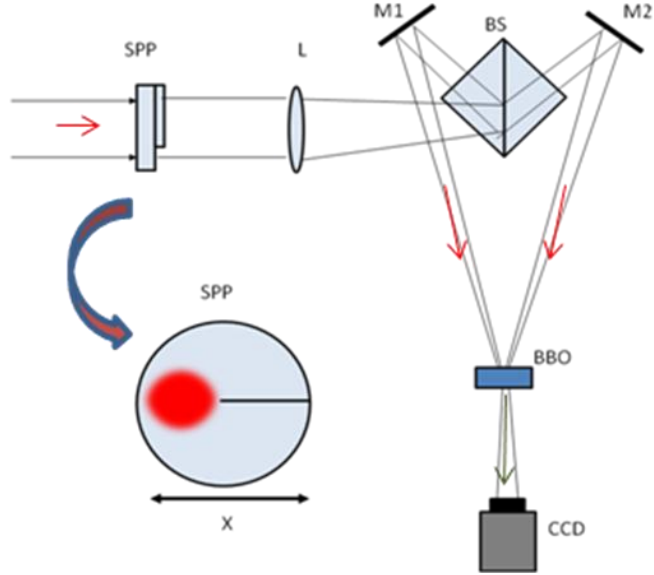

Fig. 1 Experimental set-up of noncollinear second harmonic generation. The collimated beam impinges on the spiral phase plate, then is split and focused on the nonlinear crystal with relative angles of +/- 3.5° (total 7°). In the bottom-left corner is represented the scan of the Gaussian beam performed by moving the SPP in the horizontal direction.

The generated vortex beam is split in two beams of about the same intensity and reflected by the mirrors M1 and M2; the tightly focused beams intersect in the focus region with an angle of 7° with respect to one another, on the BBO crystal of 1mm length. The SPP is mounted on a XY translational stage in order to perform a full scan of the device. Therefore the phase plate was displaced from the beam's propagation axis both in the horizontal and vertical directions, in a 8 mm x 8 mm range, with a 100 μm step. The displacement of the SPP causes the horizontal movement (besides the vertical one) of the mixed screw-edge dislocations, enclosed in the fields reflected by the mirrors M1 and M2, in opposite directions; since in this way one of the FF fields experiences an even number of reflections while the other an odd number, fields with opposite values of OAM impinge on the input plane of the crystal.
At the output plane of the crystal the two fundamental (FF) fields come out with their collinear SH signals with the same angle which they had at the input plane, while the outgoing noncollinear SH signal propagates along the bisector of the total incidence angle. The latter is finally filtered and collected by a CCD camera about 30 cm far from the BBO crystal. The noncollinear second harmonic generation occurs under type I phase matching conditions.

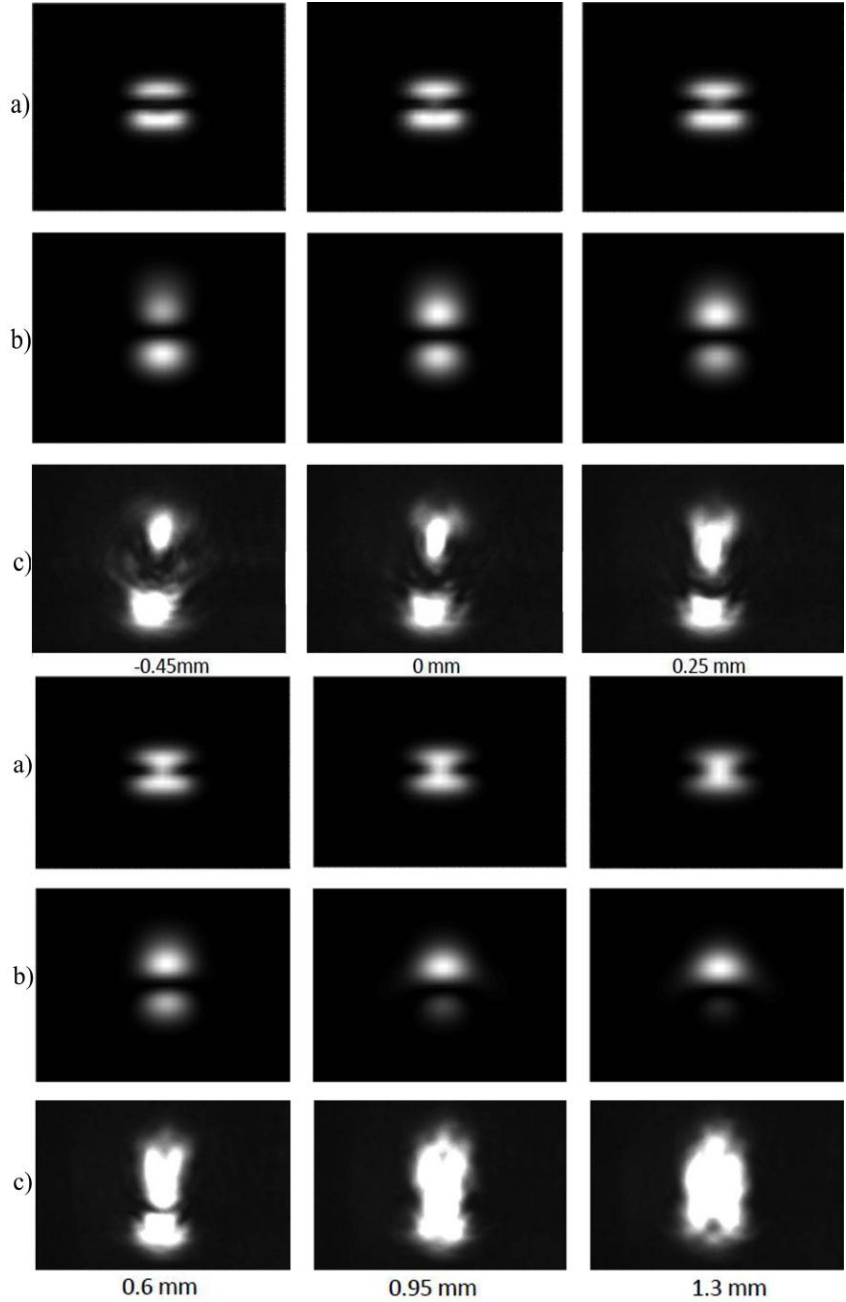

Figure 2. Noncollinear SHG signal obtained by displacing the SPP along x axis at y = 0, for six different x displacement values. a) Numerically simulated near field; b) numerically simulated far field; c) experimental results.

An example of experimental results is shown in figure 2, where the horizontal displacement of the SPP gives rise to a SH field with different configurations. The experimental results (bottom row) come from the horizontal scan in the central area of the SPP, i.e. for y=0. The upper row is the near field distribution calculated numerically, and the middle row is the far field SHG distribution (see next section). We attribute the slight difference among experiments and numerical calculations to the finite size of the height anomaly of SPPs.

At the output plane of the crystal, we expect a SH field which carries zero OAM, namely the sum of the incident angular momenta [15, 16].

*3.2 Numerical simulations*

Numerical results are obtained by modeling the input fields by using Eq. (2) both for the on-axis and the off-axis case. The coefficients $C_{l,p}$ were calculated numerically using a 200x200 mesh and integrating over a spatial range equal to 5 times the beam waist.

The near field shape of the intensity profile of the FF beams are shown in Figs. 3 (a) and (b), when the input Gaussian beam is located at the center of the SPP, thus generating an OAM per photon of 1/2.

Considering a phase matched interaction, neglecting walk-off and under paraxial approximation [11,15], the near field noncollinear SH intensity is given by

$$I^{2\omega} \approx |E_1^{\omega}(r,\vartheta,z_0)E_2^{\omega}(r,\vartheta,z_0)|^2 , \qquad (7)$$

being $E_1$ and $E_2$ the two near field transverse distribution of the input beams. An example of SH's near field intensity distribution is given in fig.3 (c), for the input conditions of figs 3a and 3b.

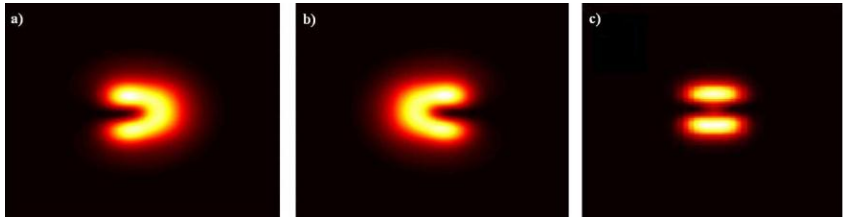

Figure 3: Numerically estimated near field intensity profiles of the FF beams, (a) and (b), and of the noncollinear SH beam; in FF beams the singularity shifts in opposite directions. The SH beam's profile carries the singularities of both the FF beams, maintaining their original orientation.

The phase of the FF and SH fields are shown in fig. 4, where the presence of phase singularities is evident. It can be noticed that the SH signal (fig. 4c) carries both the phase singularities of the incident fields, but the helical phase winds in opposite directions.

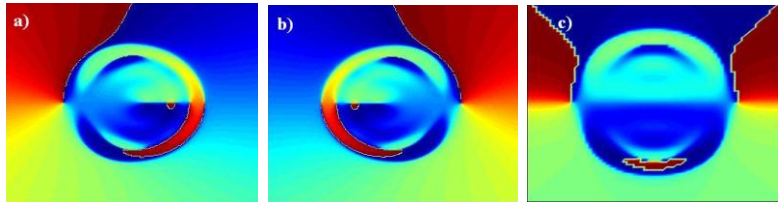

Figure 4: Phase plots of the electric field for the FFs, a) and b), and for the SH (c), corresponding to the configuration of fields given in fig. 3. The phase ranges from 0 (blue) to $2\pi$ (red).

This result, which comes out from the simplistic model adopted [11] is not surprising. In fact in the nonlinear interaction the phases of the FFs are added, thus the singularities compose in order to cancel each other.

In order to perform a comparison among experimental result and modeling, far field calculations of the SH signal have been carried out; in this way, still under paraxial approximation, the intensity of the second harmonic field in a plane transverse to the noncollinear SH field's propagation direction is given by :

$$I^{2\omega}(x,y,z) \approx \left| \int dr' E_1^{\omega}(x',y',z_0) E_2^{\omega}(x',y',z_0) \exp[i\frac{k}{z}(x'x+y'y)] \right|^2 \qquad (8)$$

where $(x', y')$ are the transverse coordinates on the plane $z = z_0$. Some results are shown in fig.2 (middle row). In figure 5 is reported a frame of a movie showing the far field of the second harmonic generated signal, together with its corresponding input fields, as a function of the phase plate's displacement.

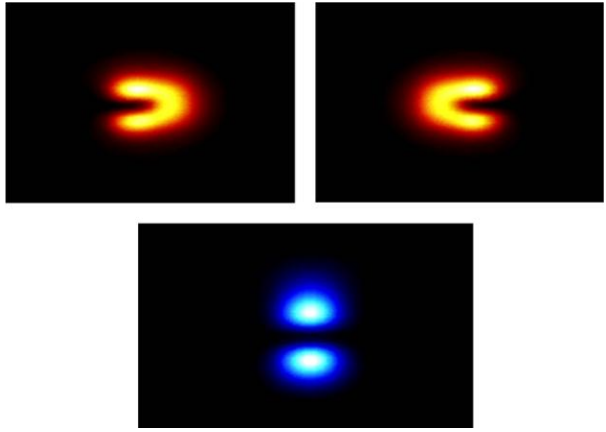

Figure 5: Frame of the far field movie of the SHG signal as a function of the SPP's displacement; the above frame corresponds to zero displacement (x = 0).

### 3.3 OAM calculations

We used both Eqs. (5) and (6) to estimate numerically the orbital angular momentum of the fields. The orbital angular momentum of the SH signal was numerically calculated for each step of the displacement by using Eq. (5). As expected, the OAM of the generated field is conserved [11], producing a beam with an angular momentum equal to the sum of the individual OAMs of the two input beams: as a consequence $\ell_{SH}$ is found to be always 0. Obviously, since the dislocation is not purely a screw type, in the generated field only the screw components of the singularity cancel, while the edge dislocations of both fields, which do not possess a direction, remain unaffected, thus creating the nodal lines that can be seen in the figure above.

The behavior of the OAM of the SH as a function of the SPP's displacement is presented in fig. 6 together with the angular momentum of the FFs. As pointed out before the value of the OAM of the SH is always zero whichever is the displacement, namely regardless of the FF's momentum.

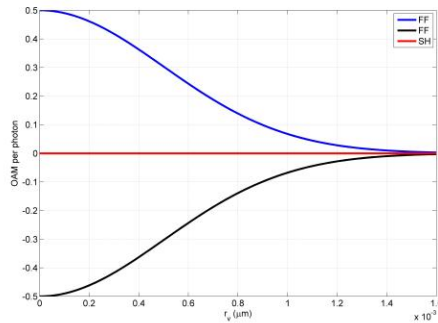

Figure 6. Orbital angular momentum of pump and second harmonic fields with the SPP's displacement. As the device is moved from the beam's axis both the FF beams (blue and black lines) reduce the absolute value of their OAM from the initial values of ± 0.5. SH's OAM (red line) is always the sum of the other two, that is zero.

The zero OAM of the SH field can be also put in evidence by calculating the Poynting vector; fig. 7 shows the Poynting vector of the SH field obtained when the displacement of the SPP is zero (as shown in fig. 3c): it can be seen that arrows are always directed along the same direction, thus the Poynting vector does not rotate, implying a zero angular momentum.

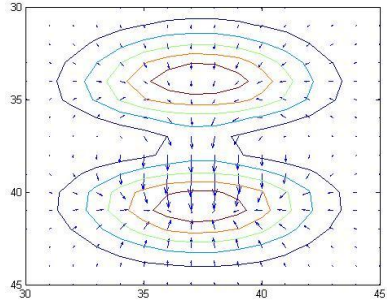

Figure 7. Transverse Poynting vector of the second harmonic field generated in the case of on-axis SPP. The absence of rotation of the arrows implies that there are no screw dislocation, therefore there is no OAM

## 4. Conclusions

This work investigated the process of noncollinear second harmonic generation involving beams carrying phase singularities generated by a half integral spiral phase plate. The focus was posed on the possibility of reaching, starting from fields with a certain orbital angular momentum, different spatial configurations, all characterized by the absence of OAM, simply by moving the SPP from the beam's axis.

Experimental results were obtained by performing a transverse scan of the SPP and by letting interact fields with opposite angular momentum. These results were then successfully compared to numerical simulations and it was found that the OAM of the generated second harmonic field is always zero and is independent on the displacement of the SPP. Furthermore the transverse component of the Poynting vector of the SH beam was calculated in order to evidence that the energy of the considered field does not circulate, confirming then the absence of the orbital angular momentum in all the spatial configuration obtained.

### Aknowledgement

We thank J.P. Woerdman and Eric Eliel at Huygens Laboratory, Leiden, for producing and providing the spiral phase plates used in our work.